\documentstyle[prd,psfig,preprint,aps]{revtex}
\begin{document}
\begin{titlepage}
\pagestyle{empty}
\begin{flushright} 
{BROWN-HET-1059} \\
{August 1996}
\end{flushright}
\begin{center} 
{\Large \bf D-DIMENSIONAL RADIATIVE PLASMA: A KINETIC APPROACH}\\
\renewcommand{\thefootnote}{\alph{footnote}}
{\small A.Maia Jr.$^{1,2,}$\footnote
{e-mail:maia@het.brown.edu} 
and J. A. S. Lima$^{2,3,}$\footnote{e-mail:limajas@fisica1.dfte.ufrn.br}}
\end{center}
\begin{quote}
{\small $^1$ Instituto de Matem\'atica - UNICAMP
13.081-970 - Campinas - S.P. - Brazil\\   
$^2$ Department of Physics, Brown University, 
Providence, RI 02912, USA \\
$^3$ Departamento de F\'{\i}sica, 
 CP 1641, Universidade 
$^{ }$ $^{ }$ 
Federal do Rio Grande do Norte, 
     59072 - 970, Natal, RN, Brazil.}
\end{quote} 
\begin{abstract}
\noindent
\baselineskip=20pt
The covariant kinetic approach for the radiative plasma, a mixture 
of a relativistic moving gas plus radiation quanta (photons, neutrinos, or gravitons)  is generalized to D spatial dimensions. The operational and physical meaning
of Eckart's temperature is reexamined and the D-dimensional expressions 
for the transport coefficients (heat conduction, bulk and shear viscosity) 
are explicitly evaluated to first order in the mean free time of the radiation quanta. Weinberg's conclusion
that the mixture behaves like a 
relativistic imperfect simple fluid 
(in Eckart's formulation) depends neither on the number of spatial 
dimensions nor  
on the details of the  collisional term. The case of Thomson scaterring is
studied in detail, and some consequences for higher 
dimensional cosmologies are also discussed.

\vspace{0.1cm}
PACS number(s): 95.30Jx, 95.30Tg, 98.80H
\end{abstract}
\end{titlepage}  
\section{Introduction}
 In many astrophysical and cosmological problems, the matter-radiation content is modeled by a so-called radiative plasma, a two component fluid consisting of some material medium (perfect fluid, relativistic gas, ionized hydrogen atoms, etc.) plus radiat
ion quanta (photons, neutrinos, or gravitons). In the simplest situation, the material 
medium is assumed to be locally in thermal equilibrium with itself (very short mean free times and mean free paths), whereas the massless component has a finite mean free time $\tau$ . However, even in this case, an idealized perfect fluid description of 
this mixture is inappropriate, and a deeper insight is provided by a kinetic theory approach. In particular, since the 
radiation quanta are out of equilibrium with the material component, all 
transport properties of this system, to  first order of approximation, depend exclusively on the mean free time of radiation. 

The first self-consistent
application of the relativistic kinetic theory to this system is due to 
Thomas\cite{TH 30}, who computed  
the correct values of the heat and shear 
viscosity transport coefficients.
Later on, by considering the concept of local equilibrium temperature, 
Weinberg\cite{W 71} 
obtained the bulk viscosity coefficient, thereby showing that such a mixture
behaves like an imperfect relativistic simple fluid in the framework of the hydrodynamical formulation for dissipative processes 
developed by Eckart\cite{E 40}. Masaki\cite{MA 71}, recovered the results of Thomas using a modern relativistic approach (manifestly covariant). This work was considerably simplified by Straumann\cite{ST 76}, who also shown that  
if the Thomson scattering dominates, 
the bulk viscosity coefficient goes to zero.
More recently, by considering the $9$-moment Grad approximation method 
to solve the Boltzman equation, Schweizer\cite{SC 82}
calculated the transient transport 
coefficients appearing in the linear, causal, 
hydrodynamical formulation proposed by Israel and Stewart\cite{IS 79}. The physical and operational meaning of local equilibrium temperature (Eckart's temperature) has also been discussed by several authors\cite{W 71},\cite{SC 82},\cite{LW 90}. 

On the other hand, since multidimensional models are presently believed to 
be the natural framework to unify all fundamental interactions, it is very common nowadays to study all phenomena and the underlying physics in an arbitrary 
number of dimensions. Historically, arbitrary dimensionality (including fractal dimension) has been widely considered 
in the context of Newtonian 
gravitational theory and general relativity, quantum field theory, cosmology, 
phase transitions, renormalization group, quantum mechanics, lattice models and random walks\cite{FT 63}-\cite{KMME 95}. 
In domain of cosmology, for instance, several 
compactification mechanisms for extra dimensions,
presumably taking place in the very early universe, have been suggested to explain 
the now
observed 4-dimensional structure of spacetime\cite{CH 80}-\cite{WH 89}.
 
Given the lack of knowledge about the initial state of the universe, the simplest dynamic approach to dimensional reduction is to construct a multidimensional anisotropic 
cosmological model based on some fundamental 
theory (superstrings, Kaluza-Klein or
Einstein's gravitational theory) where the extra spatial dimensions are collapsing while the ``physical dimensions'' are expanding, and eventually, leading to 
the present day universe. Isotropization and cosmological dimensional reduction has also been considered in the framework of 
nonequilibrium higher dimensional cosmologies. Concrete models 
have been proposed in the case of bulk and shear viscosities leading to the conclusion that viscosity may work  efficiently to reduce the   number of spatial dimensions.\cite{MM 86}-\cite{WH 89}. As a matter of
fact, this is a new version of the earlier program of chaotic 
cosmology, for which the ordinary 
dissipative 
processes (viscosity, heat flow, ...etc.) were believed to be the 
mechanisms damping out any anisotropy and inhomogeneity existing in the 
primeval plasma\cite{CM 67}. 
However, even considering that the kinetic approach is more fundamental than
its related thermodynamics,
the analysis mentioned above are usually based only on macroscopic equations. 
In particular, to the best of our knowledge, the explicit form of the transport coefficients considered in such works
have not been derived from kinetic theory in D spatial dimensions. This one of the main 
aims of this
article.

It is also worth noticing that the definition 
of a radiative plasma, in principle, is not restricted
to the class of systems mentioned above. Under certain conditions, more 
exotic system as, for instance, a mixture of massive and massless modes in superstrings models could, formally, be 
interpreted as a radiative plasma in D dimensions. The reason for this 
possibility may be easily understood. In the collisional limit, two different 
temperatures are naturally defined in an radiative plasma, namely: the matter temperature($T_m$) and the effective radiation kinetic temperature($T_r$). However, to first order in the mean free time, the latter is different from the former only if the exp
ansion rate is different from zero. The expansion works to continuously pulling  radiation and matter out of thermal equilibrium,
because the components have different cooling rates. Basically, this is the mechanism accounting for the presence of the bulk viscosity in such a mixture as well as in a relativistic simple gas\cite{UI 82}. On the other hand, even in a thermodynamic setting, a gas 
of hot strings(superstrings, or Hagedorn's fireballs) may also be described as a two temperature system when we have a mixture of heavy strings with a gas of
light strings(see, for instance, \cite{KZ 87}).

{}From the above considerations, it seems interesting to analyze the transport properties of a D-dimensional radiative plasma. In this work, by extending the covariant 4-dimensional 
kinetic approach 
developed by 
Straumann\cite{ST 76}, we compute the transport coefficients for
such a system. Hopefully, the expressions
derived here may be useful to study the
effects of dissipative processes in higher dimensional cosmologies 
as well as in modeling a mixture of massive and massless 
modes in the framework of superstring theory.

The paper is structured as follows:
In the next section, the energy-momentum tensor for a D-dimensional radiative plasma is computed exactly by solving the D-dimensional Boltzmann transport equation to first order in the mean free time of
the radiation quanta. In section 3 we discuss the physical meaning of Eckart's temperature 
and obtain the transport coefficients
of heat conduction, bulk and shear viscosities, and 
in section 4, the photon and total entropy production
are discussed. In section 5, a specific model, namely, Thomson 
scattering, is
studied. We conclude, in section 6, with a discussion of our main results for higher dimensional cosmology, and in the following two appendix we proof the basic mathematical properties used in the article. In our units the physical 
constants $\hbar=k_B=c=1$, 
where $k_B$ is Boltzmann's constant and the 
D+1-dimensional signature of the spacetime metric is 
$(+,-,-,-, . . .,-)$.    

\section{D-dimensional Kinetic Approach}

\hspace{0.3in} The basic object for 
computing the energy-momentum tensor(EMT) and other macroscopic 
quantities describing a D-dimensional radiative plasma is the radiation distribution function $F(k,x)$. As usual, it is defined in such a 
way that 
$F(k,x)d^{D}xd^{D}k$ gives the number of quanta in the D-dimensional volume element located at $\bf x$, and whose D-momentum {\bf k} lie within $d^{D}k$. Now, by assuming that 
the radiation quanta (photons for the sake of simplicity\cite{CO1}) are out but close to 
equilibrium, we may expand $F(k,x)$,
to first order in mean free time, as

\begin{equation} \label{eq:1}
F= F^{(0)} + F^{(1)} \quad,
\end{equation}
where $F^{(0)}$ is the equilibrium 
distribution in the $D$-dimensional space 

\begin{equation} \label{eq:2}
F^{(0)} = {2 \over (2\pi)^D}{1 \over e^{k.U \over T} - 1}  \quad,
\end{equation}
and  the interaction part $F^{(1)}$  satisfies
$|F^{(1)}|<< F^{(0)}$.  
In the above expressions, $U^{\mu}$ is the  normalized four-velocity field ($U_{\mu}U^{\mu}=1$) of the material medium and $T \equiv T(x)$ its local temperature which, in 
equilibrium, coincides with
the uniform temperature $ T_{_{M}} $ of the material medium. 
  
The distribution function satisfies the Boltzmann transport equation which to first order may be 
written as 

\begin{equation} \label{eq:3}
k^{\mu}{\partial F^{(0)} \over \partial x^{\mu}} = L[F^{(1)}] ,
\end{equation}
where L is a linear operator which depends on the specific model
(interaction of radiation and material fluid). Following 
Strauman\cite{ST 76}, the solution of the above equation  will be discussed, as far as possible, in rather general terms. An application of these 
results for a specific model will be presented in section 5.

As is widely known, the above Eq.(\ref{eq:3}) cannot, in general, be exactly solved so that some approximation
scheme need to be implemented. One which is often 
considered is provided by the Grad's relativistic 9-moments method applied in Refs.\cite{ST 76,SC 82} to the 4-dimensional case. In the present D-dimensional 
framework it may be generalized as is explained below.

Let $G$ be the $D$-dimensional Lorentz group of the Minkowski-like                                        space-time $M$ and 
$G_{x}$ the little group associated to a timelike vector $U(x)$. This
subgroup of $G$ leaves $U(x)$ invariant. As a 
consequence $G_{x}$ is isomorphic 
to $SO(D)$, thereby
implying that  $F(k,x)$ 
is a function 
of $ \omega = k_{\mu}U^{\mu} $, $n^{\mu}$ and $x$ where

\begin{equation} 
n^{\mu} = {k^{\mu} \over \omega}-U^{\mu} \quad.
\end{equation}
Note that 
$n_{\mu}U^{\mu} = 0$ and $ n_{\mu}n^{\mu} = -1$.
As we shall see next, the key ingredient
to make the integrals on the 
hypersphere $S_{_{D-1}}$ (space-group where $G_{x}$ acts upon) 
easier to compute comes from the fact that $\omega = k_{\mu}U^{\mu}$ is 
obviously invariant under $G_{x}$ at every point $x$.  

The calculational basis of Grad's method applied to 
radiative problems is to expand the perturbative term 
$F^{(1)}(\omega,n^{\mu},x)$ with respect to $n^{\mu}$ into irreducible polynomials under the action of the little group $G_{x}$
\begin{equation} 
F^{(1)}(\omega,n^{\mu},x) = A(\omega,x) + B_{\mu}(\omega,x)n^{\mu} + 
C_{\mu \nu }(\omega,x)(n^{\mu} n^{\nu} + { 1 \over D} h^{\mu \nu}) + ...\quad,
\end{equation}
where 
\begin{equation}
h^{\mu \nu} = \eta^{\mu \nu} - U^{\mu} U^{\nu} \;, \qquad {h^{\mu}_{\mu} = D}\quad,
\end{equation}
is the projector onto the hyperplane 
normal to $U^{\mu}$, and $\eta^{{\mu}{\nu}}$ is 
the Minkowski matrix. The former object also 
satisfies the identities:
$h^{\mu}_{\nu} U^{\nu} = 0$ and 
$h^{\mu}_{\nu} n^{\nu} = n^{\mu}$. Naturally,
the vectorial and tensorial parts in the  decomposition (5) are 
defined in the rest space normal to 
$U^{\mu}$ i.e.,  
$h^{\mu}_{\nu} B^{\nu} = B^{\mu}$ and
$h^{\sigma}_{\mu}h^{\lambda}_{\nu} C_{\lambda \sigma} = C_{\mu \nu}$. In addition, since 
$n^{\mu} n_{\nu} + {1 \over D} h^{\mu}_{\nu}$ is symmetric and traceless, without loss of generality, one may assume that the same holds for $C_{\mu \nu}$, that is: 
\begin{equation}
C_{\nu \mu} = C_{\mu \nu} \;, \quad
C^{\lambda}_{\lambda} = 0 \quad.
\end{equation}

On the other hand, since $L$ behaves like a scalar 
under the action of the little group 
$G_{x}$, this means that it operates in the irreducible subspaces 
spanned by the irreducible polynomials in (5) as a multiple of the unit operator. It thus follows that 
\begin{equation}
L[F^{(1)}] = - \omega [\kappa_{0} A + \kappa_{1} B_{\mu} n^{\mu} +
\kappa_{2} C_{\mu \nu} ( n^{\mu} n^{\nu} + {1 \over D} h^{\mu \nu}) + ...] \quad,
\end{equation}
where $\kappa_{i}=\tau_{i}^{-1}$ $(i=0, 1, 2)$, is the inverse of the mean free time for the process in consideration (bulk viscosity, heat flux, shear viscosity). As usual, we assume that these quantities are functions 
solely of $\omega$ and $x$.

Using the transport equation (3) we can express the coefficients $A$,
$B^{\mu}$ and $C_{\mu \nu}$ in terms of $U^{\mu}$, $T$, and $\kappa_{i}$.
To do that we introduce a measure $d \Omega_{U}$ on the $(D-1)$-dimensional
hypersurface
\begin{equation}
S_{_{D-1}} = \{ k / \: k^{2} = 0 , \: k^{0}>0 ,  \: k.U = const  \} \quad,
\end{equation}
for which $d \Omega_{U}$ is the unique $G_{x}$-invariant measure
normalized as
\begin{equation}
\int_{S_{_{D-1}}} \! \! {d \Omega_{U}} = {2 \pi^{D/2} \over \Gamma({D \over 2})}
\equiv S_{_{D-1}} \quad,
\end{equation}
where $\Gamma$ is the gamma function. 

This normalization is easily understood in the comoving frame. Its nothing
more than the ``area'' of unit $(D-1)$-dimensional sphere. In the above 
formula we have denoted the area of $S_{_{D-1}}$ with this same symbol, but this willnot cause any misunderstanding to the reader.

Now, with respect to this measure, the irreducible polynomials are
normalized by (see Appendix 1)

\begin{equation}
{1 \over S_{_{D-1}}} \int_{S_{_{D-1}}} {\! \! n^{\mu} n^{\nu}} d\Omega_{U} = 
-{1 \over D} h^{\mu \nu} \quad,
\end{equation}

\begin{eqnarray}
\lefteqn{{1 \over S_{_{D-1}}} \int_{S_{_{D-1}}}
(n^{\mu} n^{\nu} + {1 \over D} h^{\mu \nu})
(n^{\sigma} n^{\rho} + {1 \over D} h^{\sigma \rho}) d \Omega_{U} = } 
\nonumber \\
& & {1 \over D(D+2)} (h^{\mu\nu}h^{\sigma\rho} + h^{\mu\sigma}h^{\nu\rho} +
h^{\mu\rho}h^{\nu\sigma}) - {1 \over D^{2}} h^{\mu\nu} h^{\sigma\rho} \quad,
\end{eqnarray}

Also we have the very useful result
\begin{equation}
\int_{S_{_{D-1}}} n^{\mu} n^{\nu} \ldots n^{\sigma} d \Omega_{U} = 0 \quad,
\end{equation}
for an odd number of components $n^{(s)}$ of the vector $n$.
From a straightforward (although unduly long) calculation we can get the 
coefficients $A$, $B_{\mu}$, $C_{\mu\nu}$ in 
(5) by evaluating the moments of the
Boltzmann equation (3), and using the relations (11)-(13). The results are:
\begin{equation}
A = {1 \over \kappa_{0}} {\omega \over T_{_{M}}} 
\phi' \left( {\omega \over T_{_{M}}} \right) 
\left[ {1 \over T_{_{M}}} U^{\mu} \partial_{\mu} T_{_{M}} + 
{1 \over D} U^{\mu}_{,\mu} \right] \quad, 
\end{equation}

\begin{equation}
B_{\mu} = {1 \over \kappa_{1}} {\omega \over T_{_{M}}}
\phi' \left({\omega \over T_{_{M}}} \right)
\left[ {1 \over T_{_{M}}} h^{\nu}_{\mu} \partial_{\nu} T_{_{M}} -
U^{\nu} \partial_{\nu} U_{\mu} \right]\quad,
\end{equation}

\begin{equation}
C_{\mu\nu} = -{1 \over 2\kappa_{2}} {\omega \over T_{_{M}}} 
\phi' \left({\omega \over T_{_{M}}} \right)
\left[ h^{\lambda}_{\mu} \partial_{\lambda} U_{\nu} + 
h^{\lambda}_{\nu} \partial_{\lambda} U_{\mu} -
{2 \over D} h_{\mu\nu} U^{\lambda}_{,\lambda} \right] \quad,
\end{equation}
where $\phi({\omega \over T_{_{M}}}) \equiv F^{0} (\omega,x)$, and a prime denotes its derivative.

By definition, the total energy-momentum tensor $T^{\mu\nu}$ of the mixture (matter plus
radiation) is given by
\begin{equation}
T^{\mu\nu} = T^{\mu\nu}_{M} + T^{\mu\nu}_{R} + 
\int k^{\mu} k^{\nu} F^{(1)} {d^{D} k \over k^{0}} \quad,
\end{equation}
where  $T^{\mu\nu}_{M}$ is the EMT of the 
material component, which 
from now on, will 
be described by a perfect fluid
\begin{equation}
T^{\mu\nu}_{M} = \rho_{_{M}} U^{\mu} U^{\nu} - p_{_{M}} h^{\mu\nu} \quad,
\end{equation}
and $\rho_{_{M}}$ and $p_{_{M}}$ are, respectively,
the energy density and pressure. 
$T^{\mu\nu}_{R}$
is the radiation EMT for the equilibrium state,
which may be readily evaluated to be\cite{CO2}  
\begin{equation}
T^{\mu\nu}_{R} = \int k^{\mu} k^{\nu} F^{0} {d^{D} k \over k^{0}} =
a_{_{D}} T^{D+1}_{M} \left ({ U^{\mu} U^{\nu} - {1 \over D} h^{\mu\nu}}\right) \quad,
\end{equation}
or equivalently, in the standard form
\begin{equation}
T^{\mu\nu}_{R} = \rho_{_{R}} U^{\mu} U^{\nu} - p_{_{R}} h^{\mu\nu} \quad,
\end{equation}
where
\begin{eqnarray}
\rho_{_{R}} = a_{_{D}} T^{D+1}_{M} ,\hspace {3mm} 
p_{_{R}} = {1\over D} \rho_{R}\quad, \nonumber
\end{eqnarray}
and
\begin{equation}
a_{_{D}} = {4 \over {(2 \pi)}^{D}} {\pi^{D/2} \over \Gamma ({D \over 2})} 
\Gamma(D+1) \zeta(D+1) \quad,
\end{equation}
is the $D$-dimensional radiation 
constant, and  $\zeta$ 
denotes the Riemann zeta function.

The last term (the integral) in Eq.(17) is the only out of equilibrium contribution (due to interaction between matter and radiation), and 
can properly be interpreted as a 
dissipative term to the 
equilibrium energy-momentum tensor. 
The coefficients $A$, $B_{\mu}$, $C_{\mu\nu}$ 
give rise to contributions of different tensorial 
ranks, which in the thermodynamic theory are usually named, respectively, 
as bulk viscosity, heat flux, and shear tensor. Substituting (14)-(16) in the last integral
of (17), we get the terms:

a) Scalar (bulk viscosity) 
\hspace{1.5cm}  $F^{(1)}_{bv} = A$
\begin{equation}
\int k^{\mu} k^{\nu} F^{(1)}_{bv} {d^{D} k \over k^{0}} =  
-{(D+1) a_{_{D}} T^{D+1}_{M} \over \bar{\kappa}_{0}}
(U^{\mu} U^{\nu} - {1 \over D} h^{\mu\nu})
\left({1 \over T_{_{M}}} U^{\lambda} \partial_{\lambda} T_{_{M}} + 
{1 \over D} U^{\lambda},_{\lambda}\right)
\end{equation}

b) Vectorial (heat flow)  \hspace{1.5 cm} $F^{(1)}_{hf} = B^{\mu} n_{\mu} $
\begin{equation}
\int k^{\mu} k^{\nu} F^{(1)}_{hf} {d^{D} k \over k_{0} } = 
\left({D+1 \over D} \right) {a_{_{D}} T^{D}_{M} \over \bar{\kappa}_{1}}
(h^{\mu\lambda} U^{\nu} + h^{\nu\lambda} U^{\mu})
(\partial_{\lambda} T_{_{M}} - T_{_{M}} U^{\sigma} U_{\lambda,\sigma})
\end{equation}

c) Tensorial (shear viscosity)  \hspace{1.5cm}    $F^{(1)}_{sh} = C_{\mu\nu}(n^{\mu} n^{\nu} +
{1 \over D} h^{\mu\nu})$
\begin{equation}
\int k^{\mu} k^{\nu} F^{(1)}_{sh} {d^{D} k \over k_{0}} =
{(D+1) a_{_{D}} T^{D+1}_{M} \over D(D+2) \bar{\kappa}_{2}}
h^{\mu\sigma} h^{\nu\rho}(U_{\sigma,\rho} - U_{\rho,\sigma} -
{2 \over D} h_{\sigma\rho} U^{\lambda}_{,\lambda} ) \quad.
\end{equation} 
In the above equations, 
$\bar{\kappa}_{i}$ ($i = 0, 1, 2$) are the $D$-dimensional Rosseland means for $\kappa_{i}$.
In general, for an arbitrary quantity 
$\kappa_i$ we define its $D$-dimensional 
Rosseland mean $\bar{\kappa_i}$ by
\begin{equation}
{1 \over \bar{\kappa}_i} = {\int^\infty_0 {1 \over \kappa_i} \omega^{D+1}
\phi'({\omega \over T_{_{M}}}) d\omega  \over \int^\infty_0 \omega^{D+1}
\phi'({\omega \over T_{_{M}}}) d\omega }\quad.
\end{equation}

\section{Eckart Temperature and the Transport Coefficients}

In the phenomenological theory of 
Eckart, the EMT of a relativistic dissipative simple fluid reads  
\begin{equation}
T^{\mu\nu} = \rho(T,n) U^{\mu} U^{\nu}
 - p(T,n) h^{\mu\nu} 
 + \Delta T^{\mu\nu} \quad,
\end{equation}
where the last term is the macroscopic 
description of the irreversible processes. Its D-dimensional canonical form is given by (the factor 1/D appears explicitly because the shear viscosity stress is traceless)
\begin{eqnarray}
\lefteqn{\Delta T^{\mu\nu} = \zeta_{_{D}} h^{\mu\nu} U^{\lambda}_{,\lambda} + \chi_{_{D}} (h^{\mu\lambda} U^{\nu} + 
h^{\nu\lambda} U^{\mu})
(T_{\lambda} - T U^{\sigma} U_{\lambda,\sigma})  \nonumber} \\ 
& & + \eta_{_{D}} h^{\mu\sigma} h^{\nu\rho}
(U_{\sigma,\rho} + U_{\rho,\sigma} - {2 \over D} g_{\sigma\rho} 
U^{\lambda}_{,\lambda}) \quad,
\end{eqnarray}
where $\zeta_{_{D}}$, $\chi_{_{D}}$, and $\eta_{_{D}}$ are, respectively, the bulk viscosity, heat conducting, and the shear viscosity transport coefficients. In principle, the above expression should be justified by kinetic contributions like (22)-(24). 
It is worth noticing that all
thermodynamic quantities (explicitly or implicitly) present in (26) are measured in the rest frame of $U_{\alpha}$. However, for a radiative plasma there are, in the collisional limit,  at least three 
concepts of temperature\cite{LW 90}. Since the physical features relating these temperatures concepts are independent
(to first order in $\bar{\kappa}_{0}^{-1}$)
 of the shear and heat conducting effects, for a moment, we will restrict our discussion to the isotropic case for which only bulk viscosity is relevant.
To do that, let us first introduce (from the bulk
viscosity contribution (22)) the following auxiliary quantity: 
\begin{equation}
\Pi_{_{M}} = -(D+1) a_{_{D}} T^{D+1}_{M} {1 \over \bar{\kappa}_{0}}
\left( {1 \over T_{_{M}}} U^{\lambda} \partial_{\lambda} T_{_{M}} + {1 \over D}
U^{\lambda}_{,\lambda} \right) .
\end{equation} 
In terms of $\Pi_{_{M}}$, the isotropic part of the total EMT (17) can be written as (see (22))
\begin{eqnarray}
T^{\mu\nu} = [\rho_{M} (T_{_{M}},n) + a_{_{D}} T^{D+1}_{M} + \Pi_{_{M}}]U^{\mu}U^{\nu} -
[p_{M}(T_{_{M}},n) \nonumber \\
 + {1 \over D} a_{_{D}} T^{D+1}_{M} + {1 \over D} \Pi_{_{M}}] 
h^{\mu\nu} \quad.
\end{eqnarray}
Note that by defining the effective radiation kinetic temperature:

\begin{equation}
T_{R}=T_M\left[1-\bar{\kappa}_{0}^{-1}\left({1 \over T_{_{M}}} U^{\lambda} \partial_{\lambda} T_{_{M}} + 
{1 \over D} U^{\lambda},_{\lambda}\right)\right]\quad,
\end{equation}
the above EMT assume the form
\begin{equation}
T^{\mu\nu} = \left(\rho_{{M}}(T_M, n) + a_D T_R^{D+1}\right) U^{\mu} U^{\nu} - \left(p_{M}(T_M, n) + {1 \over D}a_D T_{R}^{D+1}\right)h^{\mu\nu} \quad,
\end{equation}
which has only two standard terms, which are characteristic of the perfect fluid
description. In particular, the radiation quanta 
behaves as if they were in equilibrium at temperature $T_{R}$ (compare (19)). However, the difference in temperature between the material and radiative components give rise to an irreversible heat transference between them. This is the physical mechanism 
accounting for the bulk viscosity process in this mixture. In order to extract its analytical expression, 
we recall that the Eckart temperature $T$ is defined as being the local equilibrium temperature, which is implicitly fixed by
(see Ref.\cite{W 71})
\begin{equation}
U_{\mu}U_{\nu} T^{\mu\nu} = \rho(T,n) = \rho_{M}(T_{_{M}},n) + a_{_{D}} T^{D+1}_{M} +
\Pi_{_{M}} = \rho_{M}(T,n) + a_{_{D}} T^{D+1} .
\end {equation}
Since the equilibrium temperature $T \in 
[T_M, T_R]$, whose length $|T_R - T_M|$ is of first order 
in $\bar{\kappa}_{0}^{-1}$, we can expand $\rho(T,n)$ as a power series
\begin{equation}
\rho(T,n) = \rho(T_{_{M}},n) + \left( {\partial \rho \over \partial T}\right)_{n}
(T-T_{_{M}}) + \cdots \quad,
\end{equation}
where
\begin{equation}
\rho(T_{_{M}},n) = \rho_{M}(T_{_{M}},n) + a_{_{D}} T^{D+1}_{M} \quad.
\end{equation}
Inserting (34) into (33) and comparing with (32) we get
\begin{equation}
T = T_{_{M}} + \left({\partial\rho \over \partial T}\right)^{-1}_{n} \Pi_{_{M}} \quad.
\end{equation}
Analogously, the total pressure $p(T,n)$  can be expanded, to first order, as
\begin{equation}
p(T,n) = p(T_{_{M}},n) + \left( {\partial p \over \partial T } \right)_{n}
(T - T_{_{M}}) + \cdots \quad,
\end{equation}
or still, inserting  $p(T_{_{M}}, n)$ from (31), and using (35)
\begin{equation}
p(T,n) = p_{_{M}}(T_{_{M}},n) + {1 \over D} a_{_{D}} T^{D+1}_{M} +
\left({\partial p \over \partial \rho}\right)_{n} \Pi_{_{M}} .
\end{equation} 
With the help of (28), (32), and (37), 
the total energy-momentum tensor (29) takes 
the form
\begin{equation}
T^{\mu\nu} = \rho(T,n) U^{\mu} U^{\nu} - \left\{ p(T,n) - {D+1 \over \bar{\kappa}_{0}} a_{_{D}} T^{D+1} \left[{1 \over D}
- \left( \partial p \over \partial \rho \right)_{n} \right]^{2}U^{\lambda}_{,\lambda}\right\}
h^{\mu\nu} 
\end{equation}
where we have used the thermodynamic relation\cite{W 71}
\begin{equation}
{U^{\lambda} \partial_{\lambda} T_{_{M}} \over T_{_{M}}} = 
- \left( \partial p \over \partial \rho \right)_{n} U^{\lambda}_{,\lambda} + O(\bar{\kappa}_{0}^{-1}) 
\quad,
\end{equation}
and the fact that, to first order, $T_{_{M}}$ can be replaced by $T$ when 
multiplied by $\bar{\kappa}_{0}^{-1}$. The above EMT has exactly the Eckart's canonical isotropic form, namely:
\begin{equation}
T^{\mu\nu} = \rho(T,n) U^{\mu} U^{\nu} - [ p(T,n) - \zeta_{_{D}} \theta]
h^{\mu\nu} \quad,
\end{equation}
where $\theta = U^{\lambda}_{,\lambda}$
is the expansion parameter, and from (38) and
(40) we read
\begin{equation}
\zeta_{_{D}} = {D+1 \over \bar{\kappa}_{o}} a_{_{D}} T^{D+1} \left[{1 \over D}
- \left( \partial p \over \partial \rho \right)_{n} \right]^{2} \quad.
\end{equation}

To obtain the remaining 
coefficients it is necessary only to observe that the vectorial and tensorial contributions (23) and (24) are not modified (to the first order) when rewritten in terms of the Eckart temperature. Hence, a direct 
comparison with the canonical
form (27) give us
\begin{equation}
\chi_{_{D}} = {D+1 \over D} {a_{_{D}} T^{D} \over \bar{\kappa}_{1}} \quad,
\end{equation}
and

\begin{equation}
\eta_{_{D}} = {(D+1) a_{_{D}} T^{D+1} \over D(D+2) \bar{\kappa}_{2}} \quad.
\end{equation}

Expressions (41), (42), and (43) are the 
transport coefficients for a D-dimensional radiative plasma. In particular, 
for $D=3$, the standard results 
are recovered\cite{W 71,ST 76}. From (41) 
we see that
structureless point particles in D-dimensions have negligible bulk viscosity in the extreme relativistic regime ($p \cong {1  \over D}\rho$), as 
should be expected in physical grounds.                                                                                                                                          This is 
also a consequence 
of the fact that the D-dimensional equilibrium 
distribution of photons is preserved under expansion, or equivalently, that
the ratio $\frac{T}{\omega}$ is an adiabatic invariant in the sense of 
Ehrenfest, regardless of the number of spatial dimensions. Hence, when the
universe is radiation dominated, there is no radiative bulk viscosity in  
the D-dimensional Friedmann-Robertson-Walker(FRW) spacetime. This generalize 
the widely known 3-dimensional result\cite{W 71}.

\section{Entropy Production for a D-dimensional Radiative Plasma}

In Eckart's frame, the current of entropy
and the local entropy production  for a D-dimensional relativistic simple fluid are given by
\begin{equation}
S^{\mu} = s(T,n) U^{\mu} + {q^{\mu} \over T}
\quad,
\end{equation} 
and
\begin{equation}
\partial_{\mu} S^{\mu} = {\Pi^{2} \over \zeta_{_{D}} T} + {q_{\mu}q^{\mu} \over \chi_{_{D}} T^{2}}+ {\Pi_{\mu\nu} \Pi^{\mu\nu} \over 2\eta_{_{D}} T} \quad,
\end{equation}
where $s$ is the entropy density, and $\Pi$, $q^{\mu}$, and  $\Pi^{\mu\nu}$ are the classical dissipative stresses (bulk viscosity, heat flow and
shear viscosity):
\begin{equation}
\Pi=\zeta_{_{D}} \theta \quad,
\end{equation}
\begin{equation}
q^{\mu}= \chi_{_{D}} h^{{\mu}{\lambda}}(T_{,\lambda} - T U^{\sigma}
 U_{\lambda,\sigma}) \quad,
\end{equation}
\begin{equation}   
\Pi^{\mu\nu}=\eta_{_{D}} h^{\mu\sigma} h^{\nu\rho}
(U_{\sigma,\rho} + U_{\rho,\sigma} - {2 \over D} h_{\sigma\rho} 
U^{\lambda}_{,\lambda})\quad.
\end{equation}
In principle, the Eckart description for a D-dimensional radiative plasma will be considered complete only if   
the kinetic equations lead to (44) and (45)
in Eckart's frame. In particular, one may argue
that the above defined entropy density should be  introduced, in a consistent way, by the same kind of algorithm used to fix the  Eckart temperature. Indeed, this is exactly what happens. To  show that let us consider the kinetic expression for the  entro
py flux\cite{SC 82}
   
\begin{equation}
S^{\mu}_{R} = - {2 \over (2\pi)^{D}} \int {d^{D} k \over k^{0}}[ F\ln F -
(1+F)\ln (1+F)] \quad,
\end{equation}
where from (2), (14), (15), and  (16) the distribution function $F$ given in (5) may be written as 
\begin{eqnarray}
\lefteqn{F = \phi \left({\omega \over T_{_{M}}} \right) + 
\kappa^{-1}_{0} {\omega \over T_{_{M}}}
\phi' \left( {\omega \over T_{_{M}}} \right) \widetilde{\Pi} +
\kappa^{-1}_{1} {\omega \over T_{_{M}}} \phi' \left({\omega \over T_{_{M}}} \right)
\widetilde{q}_{\mu} n^{\mu}\nonumber}  \\ 
& & - {\kappa^{-1}_{2} \over 2} {\omega \over T_{_{M}}}
\phi' \left( {\omega \over T_{_{M}}} \right) \widetilde{\Pi}_{\mu\nu}
(n^{\mu} n^{\nu} + {1 \over D} h^{\mu\nu})\quad.
\end{eqnarray}
where
\begin{equation}
\widetilde{\Pi} = {1 \over T_{_{M}}} U^{\mu} \partial_{\mu} T_{_{M}} +
{1 \over D} U^{\mu}_{, \mu}
\end{equation}
\begin{equation}
\widetilde{q}_{\mu} = {1 \over T_{_{M}}} h^{\nu}_{\mu} \partial_{\nu} T_{_{M}} -
U^{\nu} \partial_{\nu} U_{\mu}
\end{equation}
\begin{equation}
\widetilde{\Pi}_{\mu\nu} = h^{\lambda}_{\mu}  U_{\nu,\lambda} +
h^{\lambda}_{\nu}  U_{\mu,\lambda} - 
{2 \over D} h_{\mu\nu} U^{\lambda}_{, \lambda}
\end{equation}
Inserting the above expression into (49) a straightforward, 
although lengthy, calculation yields

\begin{equation}
S^{\mu}_{R} = \left({{D+1 \over D} a_{_{D}} T^{D}_{M} + {{\Pi_M}
\over T_M}}
\right) U^{\mu} + {q^{\mu}_{M} \over T_{_{M}}}\quad,
\end{equation}
where $q^{\mu}$ has been indexed  to recall 
that it is in the matter temperature. However,
since $q^{\mu}_M$ is already of first order in 
$\bar{\kappa}_{1}^{-1}$, this means that 
${q^{\mu} \over T_{_{M}}} \cong {q^{\mu} \over T}$, and 
using the definition of $T_R$ given by (30), the above equation becomes
\begin{equation}
S^{\mu}_{R} = {D+1 \over D} a_{_{D}} T^{D}_{R} U^{\mu} + {q^{\mu} \over T}\quad.
\end{equation}
Note that the radiation entropy density is $U_{\mu}S^{\mu}_R= {D+1 \over D} a_{_{D}} T^{D}_{R}$, as should be expected in physical grounds. In particular, for $D=3$, one has $s_R \sim T^{3}_R$,
which, in the absence of heat flow, 
corresponds to the usual radiation fluid 
with $\rho_R \sim T^{4}_R$. These results are in agreement with our discussion in the earlier section. The quanta behave (in the isotropic case) as a radiation perfect fluid at temperature $T_R$ regardless of the number of dimensions.

Now, as a self-consistency check of the 
algorithm used to introduce the temperature 
of Eckart, we need only to show that the total entropy density of the mixture reduces to the equilibrium expression at the temperature T. 

The total entropy of matter plus radiation is given by 
$S^{\mu} = S^{\mu}_{M} + S^{\mu}_{R}$ where $S^{\mu}_{R}$ is given by (55)
and $S^{\mu}_{M} = s_{_{M}}(T_{_{M}},n) U^{\mu}$ is the entropy flux of matter. 
Since the material component is in equilibrium at temperature $T_M$, the total entropy density can be written as (see (54))  
\begin{equation}
U_{\mu}S^{\mu} = s_{M}(T_{_{M}},n) + {D+1 \over D} a_{_{D}} T^{D}_{M} +
{\Pi_{_{M}} \over T_{_{M}}} \quad, 
\end{equation} 
where $s_{_{M}}(T_{_{M}}, n)$ is the entropy density of the material medium.
By definition of 
Eckart's frame, the right hand side of the above 
equation should be $s(T,n)$. In fact, 
expanding $s(T,n)$ in power series
\begin{equation}
s(T,n) = s(T_{_{M}},n) + \left({\partial s \over \partial T}
\right)_{n} (T-T_{_{M}}) + 
o(\bar{\kappa}_{0}^{-2}) \quad,
\end{equation}
and using (35), the above expression (to first
order) becomes
\begin{equation}
s(T_{_{M}},n) = s(T,n) - \left({\partial s \over \partial \rho}
\right)_{n} \Pi_{_{M}} \quad,
\end{equation}
where
\begin{equation}
s(T_{_{M}}, n)= s_{_{M}}(T_{_{M}},n) + {D+1 \over D} a_{_{D}} T^{D}_{M} \quad.
\end{equation}

Inserting (58) into (56), we can rewrite the 
entropy density as
\begin{equation}
U_{\mu}S^{\mu} = s(T,n) - \left({\partial s \over \partial \rho}
\right)_{n} \Pi_{_{M}} + {\Pi_{_{M}} \over T}  \quad,
\end{equation}
where we have used that 
${\Pi_{_{M}} \over T_{_{M}}} \cong {\Pi_{_{M}} \over T} $
to first order in $\bar{\tau}_{0} = \bar{\kappa}_{0}^{-1}$.
Finally, by considering the standard relation $\left({\partial s \over \partial
\rho }\right)_{n} = {1 \over T}$, we obtain 
\begin{equation}
U_{\mu}S^{\mu} = s(T,n)\quad.
\end{equation}
Thus, in Eckart's temperature, the total entropy four-vector of a radiative
plasma, assume the same form usually applied for a relativistic simple fluid (see (44)).

Now, we proceed to calculate the photon entropy production, namely,
$\partial_{\mu} S^{\mu}_{R}$. From (49) we have
\begin{equation}
\partial_{\mu} S^{\mu}_{R} = - {2 \over (2 \pi)^{D}} \int {d^{D} k \over
k^{0}} k^{\mu} \partial_{\mu} [F \ln F - (1+F)\ln(1+F)] \quad.
\end{equation}
Using the Boltzmann equation (3) a straightforward calculation leads to
\begin{equation}
\partial_{\mu} S^{\mu}_{R} = {2 \over (2\pi)^{D}} \int {d^{D} k \over k^{0}}
{\omega \over T} L[F^{(1)}] + {2 \over (2\pi)^{D}} \int {d^{D}k \over k^{0}}
{F^{(1)} \over \Delta^{(0)} F^{(0)}} L[F^{(1)}] \quad,
\end{equation}
where $\Delta^{(0)} = 1 + F^{(0)}$.
Using (50) these integrals are readily evaluated
\begin{equation}
\partial_{\mu} S^{\mu}_{R} = (D+1) a_{_{D}} T^{D} \left[ \widetilde{\Pi} +
{\widetilde{\Pi}^{2} \over \bar{\kappa}_{0}} + {\tilde{q}_{\mu} \tilde{q}^{\mu}
\over D \bar{\kappa}_{1}} + {1 \over 2D(D+2)} {\widetilde{\Pi}_{\mu\nu}
\widetilde{\Pi}^{\mu\nu} \over \bar{\kappa}_{2}} \right] \quad,
\end{equation}
where the linear term came from the first integral in (63). As it will be seen, 
if we include the contribution from the material medium to the 
total entropy production, we
render the total linear term zero as a consequence of the conservation of the 
total energy-momentum tensor(such approach has been applied by 
Schweizer\cite{SC 82} for the
3-dimensional case). In order to do this we first introduce the notation
\begin{equation}
{\omega \over T} = k^{\nu} {U_{\nu} \over T} \equiv k^{\nu} \beta_{\nu} \quad.
\end{equation}
and using the Boltzmann equation (3), (65), and (19), the first integral in 
(63) can be evaluated leading to
\begin{equation}
\partial_{\mu} S^{\mu}_{R}  = \beta_{\nu} T^{\mu\nu}_{R,\mu} + QP \quad,
\end{equation}
where QP stands for the second integral in (63) which gives rise to the 
quadratic part in (64).
Similarly, for the material medium one has
\begin{equation}
\partial_{\mu} S^{\mu}_{M} = \beta_{\nu} T^{\mu\nu}_{M,\mu} \quad.
\end{equation}

Adding (66) and (67), and using the conservation of the total energy-momentum tensor, $T^{\mu\nu} = T^{\mu\nu}_{R} + T^{\mu\nu}_{M}$, we get
\begin{eqnarray}
\partial_{\mu} S^{\mu} = QP = (D+1) a_{_{D}} T^{D} \left[{\widetilde{\Pi}^{2}
\over \bar{\kappa}_{0}} + {\tilde{q}_{\mu} \tilde{q}^{\mu} \over
D\bar{\kappa}_{1}} + {1 \over 2D(D+2)} { \widetilde{\Pi}_{\mu\nu}
\widetilde{\Pi}^{\mu\nu} \over \bar{\kappa}_{2}} \right] \quad.
\end{eqnarray}
With a bit more algebra one can prove that
\begin{equation}
(D+1) a_{_{D}} T^{D} {\widetilde{\Pi}^{2} \over \bar{\kappa}_{0}} = {1 \over T}
{\Pi_{_{M}}^{2} \over \zeta_{_{D}}} \quad,
\end{equation}
\begin{equation}
{D+1 \over D} a_{_{D}} T^{D} {\tilde{q}_{\mu} \tilde{q}^{\mu} \over
\bar{\kappa}_{1}} = {1 \over \chi_{_{D}} T^{2}} q_{\mu} q^{\mu} \quad,
\end{equation}
\begin{equation}
{(D+1) a_{_{D}} T^{D} \over 2D(D+2) \bar{\kappa}_{2}} \widetilde{\Pi}_{\mu\nu}
\widetilde{\Pi}^{\mu\nu} = {\Pi_{\mu\nu} \Pi^{\mu\nu} \over 
2 \eta_{_{D}} T} \quad,
\end{equation}
where $\Pi_{\mu\nu} = \eta_{_{D}} \widetilde{\Pi}_{\mu\nu}$.
Finally, using (69)-(71) we can write (68) in the expected final form
\begin{equation}
\partial_{\mu} S^{\mu} = {1 \over T} {\Pi^{2} \over \zeta_{_{D}}} +
{1 \over \chi_{_{D}} T^{2}} q_{\mu}q^{\mu} + {1 \over 2\eta_{_{D}} T}
\Pi_{\mu\nu} \Pi^{\mu\nu} \quad.
\end{equation}

\hspace{0.3in}

\section{Model with Thomson Scattering}

In D dimensions, the transport equation for the case of a nondegenerated
material medium and temperature not too high, may be written in the form (see, for example,
references [5],[6], and [14])
\begin{eqnarray}
k^{\mu} \partial_{\mu} F =
- \omega N \sigma_{a}
(1- e^{- \omega /T})(F - F^{(0)}) \nonumber \\
 - \omega N \sigma_{s}
\left[F-\int_{S_{_{D-1}}} p(n,n') F(\omega,n')  
d \Omega^{'}_{U}\right]\quad.
\end{eqnarray}
In this expression, the quantities $\sigma_{a}(\omega)$ and 
$\sigma_{s}(\omega)$ appearing in the
linearized collisional term are, respectively, the absorption and scattering 
cross sections in $D$ spatial 
dimensions, $N$ is the number of scatters and we have considered coherent scattering only. We do not evaluate 
them here, since for our purposes 
this is not relevant. The phase 
function 
$p(n,n')$ is given by relation
\begin{equation}
d\sigma_{s}(n.n') = p(n,n') \sigma_{s} d\Omega^{'}_{U} \, .
\end{equation}

If we consider only Thomson scattering, $(\sigma_{s} = \sigma_{_{Th}})$,
 we have in $D$-dimensions (see Appendix B)
\begin{equation}
p(n,n') = {1 \over S_{_{D-1}}} \left[ 1 + {D \over (D-1)^{2}}
(n^{\mu} n^{\nu} + {1 \over D} h^{\mu\nu})
(n'_{\mu} n'_{\nu} + {1 \over D} h_{\mu\nu}) \right] \quad, 
\end{equation}
which reduces to the usual Thomson formula for $D = 3$ (see \cite{Car 72,Har 88}).

In order to compute the integral in (73), we consider the total distribution function 
as given by (1), (2), and (5), namely
\begin{equation}
F = F^{(0)} + A + B_{\mu} n^{\mu} + C_{\mu\nu} ( n^{\mu} n^{\nu} +
{1 \over D} h^{\mu\nu} ) \quad.
\end{equation}

Inserting (75) and (76) into (73), and evaluating the integral, we get 
\begin{eqnarray}
\lefteqn{k^{\mu} \partial_{\mu} F = - \omega \left\{ N \sigma_{a}(\omega)
(1 - e^{-\omega/T}) A + N \left[ \sigma_{a}(\omega) (1 - e^{-\omega /T} )
+\sigma_{_{Th}} \right] B_{\mu} n^{\mu} \right. \nonumber }\\ & &
\left. +N \left[ \sigma_{a}(\omega) (1 - e^{-\omega/T}) + {D(D^{2} -3) \over
(D-1)^{2} (D+2)} \sigma_{_{Th}} \right] C_{\mu\nu}
(n^{\mu} n^{\nu} + {1 \over D} h^{\mu\nu} ) \right\} \quad,
\end{eqnarray}

Comparing the above expression with equation (8) we obtain the transport 
coefficients
\begin{equation}
\kappa_{0}(\omega) = N \sigma_{a}(\omega) (1 - e^{-\omega/T}) \quad,
\end{equation}
\begin{equation}
\kappa_{1}(\omega) = N \left[ \sigma_{a}(\omega) (1 - e^{-\omega/T}) + 
\sigma_{_{Th}} \right] \quad,
\end{equation}
\begin{eqnarray}
\kappa_{2}(\omega) = N[\sigma_{a}(\omega) (1- e^{-\omega/T}) + {D(D^{2} - 3) \over (D-1)^{2} (D+2) }
\sigma_{_{Th}}]\quad.
\end{eqnarray}

Note that when the Thomson scattering dominates, $\kappa_{0} (\omega)$ is
negligible, thereby implying that there is no radiative bulk viscosity 
in this case (see (25) and (41)). Under such a condition an homogeneous and isotropic mixture of matter and radiation could expand in equilibrium. However, for an inhomogeneous and anisotropic medium, heat conduction and shear viscosity are always present
. The Rosseland means for
$\kappa_{1}$ and $\kappa_{2} $ reads
\begin{equation}
{1 \over \bar{\kappa}_{1}} ={1 \over N \sigma_{_{Th}}} \quad,
\end{equation}
\begin{equation}
{1 \over \bar{\kappa}_{2}} = {1 \over N} {(D-1)^{2}(D+2) \over
D(D-3) \sigma_{_{Th}}} \quad.
\end{equation}

It thus follows that the transport coefficients (41), (42), and (43) are given by:
\begin{equation}
\chi_{_{D}} = {D+1 \over D} {a_{_{D}} \over N \sigma_{_{Th}}} T^{D} \quad,
\end{equation}
\begin{equation}
\eta_{_{D}} = {(D+1)(D-1)^{2} \over D^{2}(D^{2} - 3)} {a_{_{D}} T^{D+1} \over
N \sigma_{_{Th}}} \quad,
\end{equation}
\begin{equation}
\zeta_{_{D}} = 0 \quad.
\end{equation}
In particular, for $D=3$, the coefficients first computed by Straumann\cite{ST 76} are recovered.

\hspace{0.3in}
\section{Conclusion}
As we have seen, kinetic theory is indeed 
a very powerful approach to describe radiative plasmas irrespective of the number of spatial dimensions. 
It provides the transport coefficients and the standard entropy production rate, which is obtained from non-equilibrium thermodynamics. In addition, through the same algorithm used to  define the Eckart temperature, the main macroscopic quantities are also consistently settled to 
have the same form of equilibrium, as required by 
the first order nonequilibrium thermodynamics.

The Eckart approach considered here is plagued 
with  serious undesirable features, like instabilities of the equilibrium 
states and superluminal velocities of the thermal and viscous signals                                                                                                                                                                                                                                                                                                                                                                                         \cite{IM 69}-\cite{HL 87}. Such shortcomings can be circumvented 
by the so-called second order thermodynamic theories as developed
by Muller\cite{IM 69} and Israel  \cite{WI 76}. In this way, it is 
interesting to reconsider the kinetic treatment 
developed here in the framework of the causal or transient 
relativistic kinetic theory\cite{IS 79}. The corrections of the extended theories, however, will be important only in regimes where 
the mean free paths are comparable to the macroscopic 
length scales\cite{DG 80}.
 
Finally, we remark that is not difficult to widen the scope of the results derived here to include general relativistic effects. As usual, one must assume that the mean free time of each process
is much smaller than any typical scale of the
gravitational field, and at the level of the 
computations, to implement the minimal
coupling, viz.; to replace the ordinary derivatives by covariant 
derivatives and the (D+1)-Minkowski matrix $\eta^{{\mu}{\nu}}$ by $g^{{\mu}{\nu}}$\cite{W 71}. Naturally, D-gravity does not 
change the form of the D-dimensional transport coefficients, which 
should be considered in the
analysis of the entropy production in high 
dimensional cosmologies, as well as in the role
played by dissipative processes on the problem of 
dimensional reduction. In particular, 
as shown in section 3, the radiative bulk viscosity 
does not play any role in a D-dimensional 
radiation-dominated FRW universe. This generalizes the well known results obtained by Weinberg in the three dimensional case\cite{W 71}. Qualitatively,   significant bulk viscosity may
take place only if the plasma 
does a transition, for instance, from
radiation to a matter-dominated phase. 
However, if only the Thomson scaterring is 
considered, the bulk viscosity also vanishes, regardless of the number of spatial dimensions.
Further considerations in these topics will
be presented in a forthcoming communication. 

\hspace{0.3in}
\section{Acknowledgment}
It is a pleasure to thank R. Abramo, R. Brandenberger, and P. C. de Oliveira for valuable suggestions. We are also grateful for the warm hospitality of the Physics
Department at Brown University. This work was partially supported by FAPESP (Sao Paulo Research Foundation-Brazil), CNPq
(Brazilian Research Agency), and by U.S. Department of
Energy under Grant No. DE-FG02-91ER-40688-Task A.

\appendix
\section{Irreducible Polynomials Normalization} 
For completeness, we now outline the proofs of formulas (11)-(13), which
are the key ingredients to perform integrals defining the usual 
thermodynamic quantities. These formulae on the normalization of the 
irreducible polynomials are frequently cited in
the case $D=3$ (see, for instance, ref(s) [4],[5] and [7]).
Nevertheless, as far as we know, there is
no simple proof of them in the general case. As mentioned earlier 
(see section 2) we make all calculations in
the comoving frame.

The hypersphere parametrization is given by                          
\[k_{1}^{2} + k_{2}^{2} + \cdots +k_{D}^{2} = \omega^{2} \]  
where
\begin{eqnarray}
k_{1} & = & \omega \cos \theta_{1} \nonumber \\
k_{2} & = & \omega \sin \theta_{1} \cos \theta_{2} \nonumber \\
k_{3} & = & \omega \sin \theta_{1} \sin \theta_{2}  
\cos \theta_{3}  \nonumber \\ 
\vdots \nonumber \\
 k_{D-1} & = & \omega \sin \theta_{1} \sin \theta_{2} \ldots 
\sin \theta_{D-2} \cos \theta_{D-1} \nonumber \\
k_{D} & = & \omega \sin \theta_{1} \sin \theta_{2} \ldots
\sin \theta_{D-2} \sin \theta_{D-1} \nonumber \quad,  
\end{eqnarray} 
with $0 \le \theta_{j} \le \pi $ for $j=1,2,\dots,D-2$ 
and $0 \le \theta_{D-1} \le 2\pi$. 
The element of volume on $S_{_{D-1}}$ is given by
\[d\Omega_{U} = \sin^{D-2} \theta_{1} \sin^{D-3} \theta_{2} \ldots
\sin^{1} \theta_{D-2} \sin^{0} \theta_{D-1} d\theta_{1} d\theta_{2}
\ldots d\theta_{D-1} \quad, \]
and the volume of the $(D-1)$-sphere is given by equation (10).

\vspace{0.5cm}
{\bf a) Proof of equation (11)}

In the comoving frame $\omega = k^{0}$ and $n^{0} = 0$, thereby implying that (11) is 
valid for $\mu=0$ and $\nu = 0,1,2,3$ (both sides are identically zero). Now consider $i \neq j$. Since in this case $h^{ij} =0$, 
we need only to prove that
$ \int_{S_{_{D-1}}} n^{i} n^{j} d\Omega_{U} = 0 $,
or the equivalent formula  
$\int_{S_{_{D-1}}} k^{i} k^{j} d \Omega_{U} = 0 $ 
(note that $k^{i} = \omega n^{i} $). Assuming, without lost of generality, that $i < j $, and using the above parametrization of the $(D-1)$-sphere, it
follows that
\begin{equation}
\int_{S_{_{D-1}}} n^{i} n^{j} d \Omega_{U} = 
A \int_{0}^{\pi} d\theta_{i} \sin^{D-i} \theta_{i} \cos \theta_{i} = 0 \quad,
\end{equation}
where $A$ is a constant which appears from a multiple integral 
involving all angular variables but $\theta_{i}$.

For $i=j$ we have from spherical symmetry
\begin{eqnarray}
D \int_{S_{_{D-1}}} n_{i}^{2} d \Omega_{U} =    \sum_{j=1}^{D} \int_{S_{_{D-1}}}
n_{j}^{2} d \Omega_{U} = \nonumber\\
\int_{S_{_{D-1}}} d \Omega_{U} \sum_{j=1}^{D}
n_{j}^{2} = \int_{S_{_{D-1}}} d \Omega_{U} = S_{_{D-1}}
\quad,
\end{eqnarray}
and since 
$h^{ii}=-1$, $i=1,2,\ldots,D$ we 
can write
\begin{equation}
\int_{S_{_{D-1}}} n_{i}^{2} d \Omega_{U} = -{1 \over D} h^{ii} S_{_{D-1}} \quad,
\end{equation}
which means that (11) is valid for $\mu,\nu = 0,1,2,3 $.

\vspace{0.5cm}
{\bf Proof of equation (12)}

By using equation (11) we get from the left hand side of (12)
\begin{eqnarray}
{1 \over S_{_{D-1}}} \int_{S_{_{D-1}}} (n^{\mu} n^{\nu} + {1 \over D} h^{\mu \nu})
(n^{\sigma} n^{\rho} + {1 \over D} h^{\sigma \rho} ) d \Omega_{U} = \nonumber\\
{1 \over S_{_{D-1}}} \int_{S_{_{D-1}}} n^{\mu} n^{\nu} n^{\sigma} n^{\rho}
d \Omega_{U} - {1 \over D^{2}} h^{\mu \nu} h^{\sigma \rho} \quad.
\end{eqnarray}
Let us now consider the integral
\begin{equation}
{1 \over S_{_{D-1}}} \int_{S_{_{D-1}}} n^{\mu} n^{\nu} n^{\sigma} n^{\rho} 
d \Omega_{U} \quad.
\end{equation}
To evaluate it we observe that $h^{\mu \nu}$ is the only second rank 
$D$-dimensional tensor at our disposal which does not depends on the vector $n^{\mu}$, and
that all permutations of $(\mu \nu \sigma \rho) $ leave us with the same expression. The only 4-rank tensor matching these requirements is
given by
$h^{\mu \nu} h^{\sigma \rho} + h^{\mu \sigma} h^{\nu \rho} +
h^{\mu \rho} h^{\nu \sigma}$.
The integral must be proportional to this tensor, and the proportionality constant should depend only on the dimension $D$.
Therefore we can write
\begin{equation}
{1 \over S_{_{D-1}}} \int_{S_{_{D-1}}} n^{\mu} n^{\nu} n^{\sigma} n^{\rho}
d \Omega_{U} = f(D) (h^{\mu \nu} h^{\sigma \rho} + h^{\mu \sigma} h^{\nu \rho}
+ h^{\mu \rho} h^{\nu \sigma} ) \quad,
\end{equation}
where $f(D)$ is the constant to be determined. 
Since we have 
tensorial equation, we can evaluate the scalar $f(D)$ in the comoving frame. By specializing the integral for
$ n^{\mu} = n^{\nu} = n^{\sigma} = n^{\rho} = n^{1}= \cos \theta_{1} $, we obtain 
\begin{equation}
f(D) = {1 \over 3 S_{_{D-1}}} \int_{S_{_{D-1}}} {n_{1}}^{4} d \Omega_{U} \quad.
\end{equation}
A straightforward (although a bit lengthy) calculation leads to
\begin{equation}
f(D) = {1 \over D(D+2)} \quad,
\end{equation}
and using (A.8), (A.6) and (A.4) one immediately obtains (12). 

\vspace{0.5cm}
{\bf Proof of equation (13)}

 In the comoving frame ($n^{0} =0$), we need to proof (13) only for the
spatial components. Suppose that we have a odd number, say $p$, of components of the vector $n$. Since $ n^{i} = k^{i}/k^{0}$ we get
\begin{equation}
\int_{S_{_{D-1}}} n^{i} n^{j} \ldots n^{l} n^{r} d \Omega_{U} =
{1 \over k_{0}^{p}} 
\int_{S_{_{D-1}}} k^{i} k^{j}
\ldots k^{l} k^{r} d \Omega_{U} \quad.
\end{equation}
The vectors
$k=(k^{0}, k^{1}, k^{2}, \ldots,k^{D})$ and 
$\overline{k} =(k^{0},-k^{1}, -k^{2}, \ldots , -k^{D})$  cover
the same $D$-dimensional sphere $S_{_{D-1}}$. 
It thus follows that
\begin{equation}
\int_{S_{_{D-1}}} k^{i} k^{j} \ldots k^{r} d \Omega_{U} = (-1)^{p}  \int_{S_{_{D-1}}} k^{i} k^{j} \ldots k^{r} d \Omega_{U} \quad.
\end{equation}
Since $p$ is odd we get
$\int_{S_{_{D-1}}} k^{i} k^{j} \ldots k^{l} k^{r} d \Omega_{U} = 0 $ and so the proof is completed.

\section{$D$-dimensional Thomson Cross Section} 
In order to evaluate the phase function $p(n,n')$ one must impose some hypothesis reflecting the symmetry of the specific scattering problem. For Thomson scattering, 
the results of references [5] and [19] are easily generalized as follows:

{\bf H1} \hspace{1cm}     $p(n,n')$ is written in terms of irreducible polynomials.

{\bf H2} \hspace{1cm}     $p(n,n')$ = $p(n',n)$

These two hypothesis imply that $p(n,n')$ is constructed out from
products of an even number of irreducible polynomials. Thus, we can write
\begin{equation}
p(n,n') = A + Bn_{\sigma}{n'}^{\sigma} + 
C(n^{\mu} n^{\nu} + {1 \over D} h^{\mu \nu})
(n'_{\mu} n'_{\nu} + {1 \over D} h^{\mu \nu} )
\quad, 
\end{equation}
where A, B, and C are constants to be determined.

{\bf H3}  \hspace{1cm} $p(n,n')$ is normalized
such that: 
\begin{equation}
\int_{S_{_{D-1}}} p(n,n') d \Omega_{U} = 1
\quad.
\end{equation}

{\bf H4} \hspace{1cm}  The mean value of $n^{\mu}$ on the sphere $S_{_{D-1}}$ is zero, that
is:

\begin{equation}
\int_{S_{_{D-1}}} p(n,n') n'^{\mu} d \Omega'_{U} = 0 \quad.
\end{equation}
Substituting (B.1) into (B.2) we get $ A = {1 \over S_{_{D-1}}}$, and from 
(B.3) we get $ B = 0 $. Hence, $p(n,n')$ reduces to
\begin{equation}
p(n,n') = {1 \over S_{_{D-1}}} + C (n^{\mu} n^{\nu} + {1 \over D} h^{\mu \nu})
(n'_{\mu} n'_{\nu} + {1 \over D} h_{\mu \nu} )
\quad.
\end{equation}

To compute the coefficient $C$, we consider the  comoving frame where
$n = (0,\vec{n}) = (0, n_{1}, n_{2}, \ldots, n_{D})$,  $n'=(0,\vec{n}')=(0,n'_{1},n'_{2}, \ldots ,n'_{D})$,
and define the angle $\theta$ by 
\begin{equation}
n^{\mu} n'_{\mu} = - n^{i} n'_{i} = \cos \theta
\quad. 
\end{equation}
Inserting (B.5) into (B.4) we get
\begin{equation}
p(n,n')={1 \over S_{_{D-1}}}[ 1 + \alpha (\cos^{2}\theta - {1 \over D})] \quad,
\end{equation}
where $ \alpha = C S_{_{D-1}}$.
The differential scattering cross section is  defined by (see Fig 1)
\begin{equation}
d \sigma = \sin^{2} \phi d \Omega_{U} =
p(n,n') \sigma_{s} d \Omega_{U}
\quad,
\end{equation}
while the total scattering cross section $ \sigma_{s}$ reads
\begin{equation}
\sigma_{s} = \int_{S_{_{D-1}}}d \sigma = {D-1 \over D} S_{_{D-1}}\quad.
\end{equation}
It thus follows from (B.6), (B.7), and (B.8) 
that
\begin{equation}
\sin^{2} \phi = {D-1 \over D} [ 1 + \alpha (\cos^{2} \theta -{1 \over D})  ] \quad.
\end{equation}
{}From Fig. 1 we get the relation between the angles ( see, for example,
reference[20] for a $3$-dimensional version of our argument)
\begin{equation}
\cos \phi = \sin \theta \cos \psi \quad.
\end{equation} 
Now, consider the sphere $S_{_{D-2}} \subset S_{_{D-1}}$, such that
$\vec{n} \perp S_{_{D-2}}$, that is, $\vec{n}$ is orthogonal to all vectors
describing $S_{_{D-2}}$.
By  denoting $<{\cal A}>$ as the mean on $S_{_{D-2}}$ of an arbitrary
quantity ${\cal A}$, namely
\begin{equation}
<{\cal A}> = {1 \over S_{_{D-2}}} \int_{S_{_{D-2}}} {\cal A} d \Omega_{D-2}
\quad,
\end{equation} 
we obtain (note that only the angle $\psi$ relies on $S_{_{D-2}}$)
\begin{equation}
< \cos^{2} \psi > = {1 \over D-1}\quad,
\end{equation}
and from (B.10) and (B.12) 
\begin{equation}
< \sin^{2} \phi > = 1 - {1 \over D-1} \sin^{2} \theta \quad.
\end{equation}
Now, taking the average on $ S_{_{D-2}}$ of equation (B.9) and using (B.13) we
get 
\begin{equation}
C= {\alpha \over S_{_{D-1}}} = {D \over (D-1)^{2} S_{_{D-1}}} \quad,
\end{equation}
and inserting (B.14) into (B.4) we find the D-dimensional 
Thomson cross 
section
\begin{equation}
p(n,n')= {1 \over S_{_{D-1}}} [ 1 + {D \over (D-1)^{2}}
(n^{\mu} n^{\nu} + {1 \over D} h^{\mu \nu})
(n'_{\mu} n'_{\nu} + {1 \over D} h^{ \mu \nu})] \quad.
\end{equation}
As expected, for $D=3$ this expression reproduces the usual Thomson formula \cite{Car 72,Har 88}
\begin{equation}
p_{Th}(n,n') = {3 \over 16 \pi}(1 + \cos^{2} \theta)\quad.
\end{equation}

\hspace{0.3in}

\vspace{5cm}

\begin{figure}
\centerline{\psfig{figure=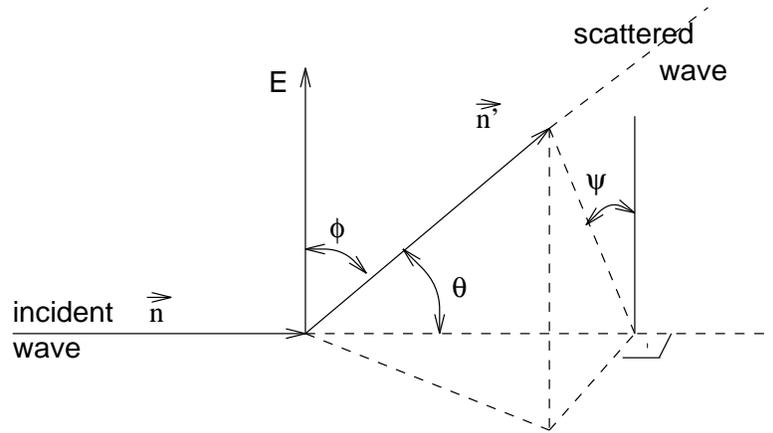,width=12cm}}
\vspace{1cm}
\caption{Thomson Scattering in $D$ dimensions. The geometry of the 
scattering is the same as in $3$ dimensions but the mean of the angle $\psi$
has been done on the $(D-2)$-dimensional sphere $S_{_{D-2}}$.}
\end{figure}


\begin{thebibliography}{99}
\bibitem{CO1} The case of neutrinos can be  
easily covered by the same formalism if one defines 
$F^{(0)} = {2 \over (2\pi)^D}\left[ {e^{k.U \over T} - \varepsilon} \right]^{-1}$, where the number $\varepsilon$ is $+1$ for bosons and $-1$ for fermions.
\bibitem{CO2} To get the above result we made the calculations in the comoving matter referential in which we have
${d^{D} k \over k^{0}} = \omega^{D-2} d \omega d\Omega_{U}$.
\end{thebibliography}
\end{document}